# Demand and Capacity Modeling for Advanced Air Mobility

Luis E. Alvarez[1], James Jones[2], Austin Bryan[3], Andrew Weinert[4]
*Massachusetts Institute of Technology Lincoln Laboratory, Lexington, MA, 02421, USA*

**Advanced Air Mobility encompasses emerging aviation technologies that transport people and cargo between local, regional, or urban locations that are currently underserved by aviation and other transportation modalities. The disruptive nature of these technologies has pushed industry, academia, and governments to devote significant investments to understand their impact on airspace risk, operational procedures, and passengers. A flexible framework was designed to assess the operational viability of these technologies and the sensitivity to a variety of assumptions. This framework is used to simulate air taxi traffic within New York City by replacing a portion of the city's taxi requests with trips taken with electric vertical takeoff and landing vehicles and evaluate the sensitivity of passenger trip time to a variety of system wide assumptions. In particular, the paper focuses on the impact of the passenger capacity, landing site vehicle capacity, and fleet size. The operation density is then compared with the current air traffic to assess operation constraints that will challenge the network UAM operations.**

## I. Introduction

The United States is reliant on robust, resilient, and efficient logistics operations. The transport of people and goods is necessary to maintain critical infrastructure, feed the population, support innovation, and sustain the domestic and global economies. To address these challenges, advanced air mobility (AAM) concepts are being developed to reduce the burden of and increasing the flexibility of logistics operations the aviation and smart city communities are developing. These concepts focus on 3-6 passenger (or equivalent) small aircraft, either manned or unmanned, traveling up to 150 miles. Urban Air Mobility (UAM) is a specific AAM concept focused on operations above and within urban environments. While UAM is a relatively new term, air travel within major urban environments is not a new concept. In the 1970s, helicopter commuting was a means of transportation between New York City and the John F. Kennedy international airport (KJFK), however safety concerns restricted operations from flying over New York City, with operations ceasing due to safety accidents [1]. Recent advancements in automation and safety technology have renewed interest in UAM concepts, especially in larger metropolitan cities [2]. Market research shows a potential for millions of operations of unmanned and manned UAM aircraft at low altitude on an annual basis. This anticipated number of on-demand air traffic is expected to stress the U.S. National Airspace System (NAS) in an unprecedented manner [3]. Unlike commercial air travel, the landing and takeoff locations (i.e., vertiports) of UAM aircraft will likely be built through private capital [4].

While there is economic and operational interest, UAM operations will still need comply with operating rules and regulations for minimizing the risk of a midair collision (MAC) between aircraft while promoting efficient use of the DISTRIBUTION STATEMENT A. Approved for public release. Distribution is unlimited.

---

[1] Technical Staff, Surveillance Systems Group, 244 Wood Street, Lexington, MA 02421.
[2] Technical Staff, Air Traffic Control Systems Group, 244 Wood Street, Lexington, MA 02421, AIAA Senior Member.
[3] Assistant Staff, Surveillance Systems Group, 244 Wood Street, Lexington, MA 02421.
[4] Associate Staff, Surveillance Systems Group, 244 Wood Street, Lexington, MA 02421, AIAA Senior Member.

This material is based upon work supported by the United States Air Force under Air Force Contract No. FA8702-15-D-0001. Any opinions, findings, conclusions or recommendations expressed in this material are those of the author(s) and do not necessarily reflect the views of the United States Air Force.

NAS. It is likely that civil authorities, such as the Federal Aviation Administration (FAA), will need to update or create new regulations tailored to UAM operations. To enable this regulatory development, government entities support testing and evaluation, and provide an initial perspective on operating procedures of UAM technologies through efforts like the NASA AAM National Campaign [5] and the FAA Urban Air Mobility Concept of Operations [6]. In alignment with these initiatives, several entities have been implementing the UAM vision by engaging in trial operations, releasing operational concepts, and studying vehicle design requirements [7], [8], [9], [10], and [11]. Although the majority of concepts envision the use of electric vertical takeoff and landing (eVTOL) aircraft, current operations make use of existing helicopters for testing procedures and market feasibility [12].

The FAA needs to understand how UAM operations can be integrated safely and efficiently analyses along with developing roadmaps will enhance decision making and the research will highlight the anticipated needs of the FAA to support further integration of UAM air transportation operations in and across metropolitan areas including suburbs and exurbs. Challenges facing UAM operations range from environment impacts to the underlying population (e.g., noise impacts), weather shutdowns, emergency landing site availability, operating flight rules, airspace capacity, and landing and takeoff locations [13]. For the design of robust operations, government and industry must understand complex infrastructure interdependences of transportation network design, landing and takeoff site capacity (vertipads), schedule assurance, and maximum sustainable throughput of the system.

In response, we developed a framework for assessing large scale, resource constrained operations of UAM. Unlike prior research which has focused on optimization of unlimited demand or resources, mode-choice, and random assignment of mission origin-destination (OD) based on census data to assess the impacts of UAM [14], [15], [16], [17], [18], the framework described in this paper approaches the problem as a parametric Monte Carlo sweep. In particular this framework asserts resource constraints, such as passenger capacity, heterogenous aircraft dynamics, vehicle availability, and vertiport vehicle landing site availability while utilizing a fractional replacement of ground for-hire vehicle data as a surrogate for OD pairs. By adopting this demand modeling strategy, there is an implicit assumption that UAM trips would have competitive pricing to ground transportation.

We assume UAM operations will be, at first, contained within well-defined corridors where traditional air traffic control (ATC) would not provide separation services. Instead, a management system similar to a UAS Traffic Management (UTM) system will provide tactical and strategic separation services [19]. Aircraft within these corridors would operate under visual flight rules (VFR) or similar [20]. The primary contributions of this paper are the design of the UAM modeling framework and demonstration of leveraging the framework to assess and characterize operational challenges of potential UAM operations in in the New York City metropolitan region. Specifically, the paper provides an insight on the sensitivity of delay to fleet size, vehicle capacity, vertiports and vertipads. The paper also analyzes the operations and compares it with non-UAM traffic operating in the airspace to understand the conflicts and challenges these operations will encounter. The rest of the paper is organized as follows. Section II defines the study scope; Section III describes the methodology being followed in the framework development. Section III defines experimental design of our analyses and accompanying results. Last, Section IV provides a summary of our study and provides insights into challenges that must be address by the UAM community.

## II. Study Scope

This paper focuses on bounding the UAM airspace requirements from an operations perspective. In particular we aim to answer the following questions:

1. What is the magnitude of UAM operations that should be expected if UAM directly replaces taxi services and at what level of replacement would stress the airspace?
2. How do fleet characteristics impact the delay experienced by passengers and flights as well as the number of operations required to meet demand?
3. How do operations and infrastructure assumptions affect operational throughput, and the efficiency of operations?
4. What combination of fleet size, vehicle capacity, vertiport capacity could reduce the travel delays of UAM operations to be competitive with ground for-hire vehicles?
5. What traffic density would be experienced in the corridor network and what interactions should be expected with non-UAM traffic?

### A. Assumed UAM Operations and CONOPS

The UAM corridor is designated airspace where aircraft must satisfy specific rules, procedures, and performance requirements. These requirements can either be operational (e.g. aircraft performance envelope, DAA equipage) or participatory (e.g. flight intent sharing, deconfliction within a specific corridor). Specifically, the FAA UAM CONOPS states that "the performance and participation requirements of UAM Corridors may vary between UAM Corridors," and that "any operator that meets the UAM Corridor performance and participation requirements may operate in, or cross, the UAM corridor." The FAA expects that access to the UAM corridor will be dependent on aircraft performance, which will likely limit operations to specific types of aircraft. The FAA assumes that aircraft within UTM corridors will primarily by rotorcraft and UAM-specific aircraft; while fixed-wing aircraft (either single or multi-engine) and UTM (e.g. smaller UAS) aircraft will likely just cross (transition through) the UAM corridors.

The FAA UAM CONOPS also emphasizes that UAM operations are distinct from UTM and ATM operations. UTM operations would be supported by Unmanned Aircraft System Service Suppliers (USS), and UAM operations by a network of Provider of Services for UAM (PSU). ATC would provide additional oversight by setting the availability of a corridor, provide advisories regarding UAM operations to other aircraft, and respond to UAM off-nominal operations. A key assumption is that aircraft, under nominal conditions, in the UAM corridor are not employing ADS-B out or transponders to communicate aircraft identification and location information. This information would be provided through the PSU. This is notably different than traditional helicopter route operations, which permit the use of ADS-B out and transponder during operations for specific airspace classes. Outside of UAM corridors, operations adhere to relevant ATM and UTM rules based on operation type, airspace class, and altitude. Thus, it is possible for a region to have its airspace allocated amongst UAM, UTM, and ATM operations.

Regarding the interaction between aircraft, the authors' interpretation of the FAA UAM CONOPS is that traditional manned aircraft would detect and surveil aircraft in the UAM corridor either through visual acquisition or a service provided by a PSU. If traditional aircraft do not cross a UAM corridor, there would be no requirement for traditional manned aircraft to be serviced by a PSU; thus we can only assume that visual acquisition is the only consistent means of surveillance. We assume any aircraft that operates within and transitions through a UAM corridor would be required to receive PSU services. These implications about airspace design are assessed further in this paper.

## III. Methodology

The UAM framework is composed of a demand model, a scheduler, and a discrete-event simulator. A demand model built from existing taxi demand drives the traffic requests for UAM. A greedy scheduler assigns aircraft to the demand based on the state of the closest available aircraft. A discrete event simulator then ingests the scheduled traffic and creates a sequence of events that are adjusted as events occurs over the course of the simulation. We leveraged the FAA UAM CONOPS v1.0 [6] when developing assumptions about UAM operations and how other aircraft can interact with these operations.

**B. Modeling Demand for Urban Air Mobility**

Using the New York Taxi database, a demand model was developed to capture the temporal and spatial variations of potential UAM demand. The New York Taxi database contains requests from 2009 to 2019 for yellow cab, green cab, and vehicle for hire (e.g., limousine, Uber, Lyft) services with a pickup and drop-off locations in the metropolitan region of New York City (NYC) [21]. For this paper the vehicle-for-hire (e.g., Uber, Lyft) data was ignored due to its inconsistent quantity of pickup and drop-off locations. Due to privacy concerns after 2016 the database does not support exact pickup and drop-off locations and instead assigns the trip to regions defined by neighborhoods. For the purpose of this paper, we use the 2016 database as it provides us with higher resolution information for pickup and drop-off. Although the underlying data for this model is ground-based, it serves as surrogate for flight demand as it includes differences in demand based on the time of the day, and day of the week. To simplify data processing, 2016 data is sorted into the seven weekdays and the number of requests are grouped based on the closest discretized bin to its original location. The demand model then uses a Poisson distribution to represent the statistical likelihood of a UAM request given the percentage of mode replacement selected. Literature has commonly used Poisson distributions to represent demand from a variety of resources as it has a well-known predictable output and reduced computational complexity [22].

The NYC metropolitan region is discretized into 100 $m^2$ areas to create sets of 360x720 grid worlds. The spatial relation between a request's OD pair is preserved by grouping the distributions based on the origination. Thus, for each hour of the day a 360x720 origin grid is created where each bin contains an equivalent 360x720 grid of possible destinations, for a total of ~67 billion pairs for each hour. The bins of a destination grid contain the number of requests observed in the time scale desired (represented by λ). This lambda parameter is then used in simulation to sample from a Poisson distribution for the specified time span desired at a specified destination bin. The sum of all random samples is then attributed to the origination bin. Afterwards a probability is assigned that this traffic will use UAM as a transport mode based on a desired percentage replacement. To reduce the demand model run time, the origin and destination grid are represented by a sparse matrix where non-zero entries are those where the NYC database shows a request existed. This representation removes grid bins over water and regions outside of the NYC metropolitan database.

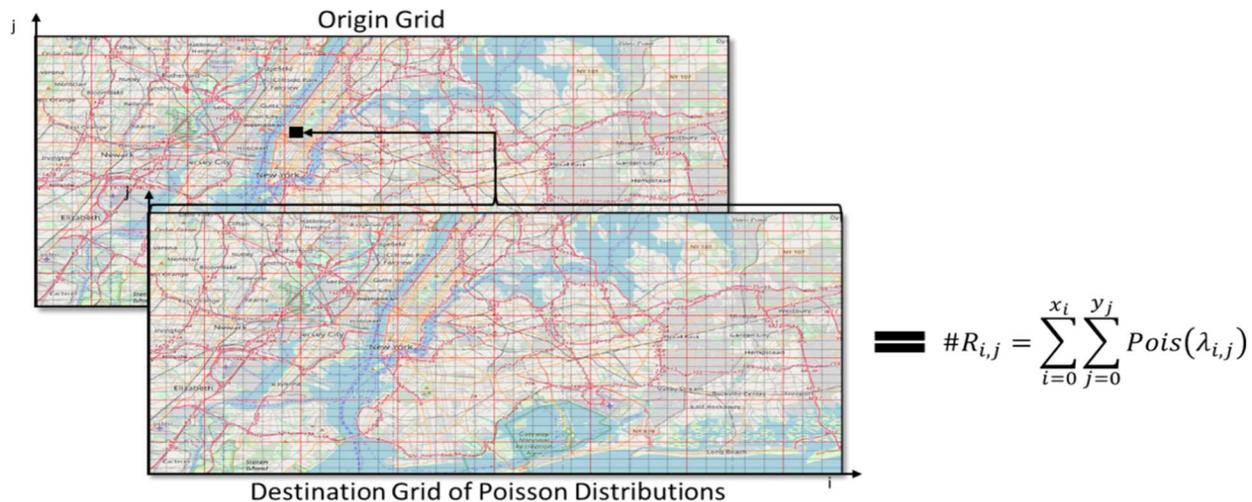

Figure 1. Mapping of origin-destination pairs by representing requested destination as a grid world with a set of Poisson distributions and attributing the cumulative requests to the origin grid world.

The symmetry in origin and destination grid allows each bin to be assigned to the nearest vertiport separate from the demand processing. During simulation run time, the demand model expects a specified time window, starting day of the week, and the time scale desired for the output. For example, the traffic associated with an OD pair can be defined as separate events for every minute of simulation or as single events occurring in an hour. At run time, the demand model collects the total number of requests for an OD pair with its associated time stamps. This information is then recorded and sent to the scheduling module as a set of events.

C. Vertiport Location and Network Selection

To determine the OD pairs of operations a set of vertiport locations is established. A similar approach to [14] was used to establish the vertiport locations, where a full year of demand is assessed with K-means clustering while considering existing heliport locations, as well as the maximum distance the passenger would have to travel. From this evaluation, we determined 29 candidate vertiport locations that are at existing heliports, open fields, and are within 1 mile of a centroid defined by the optimal K-means cluster size. **Figure 2** shows the locations of the vertiport locations and their associated catchment area. Since the taxi database is representative of NYC medallion vehicles, the vertiports in New Jersey were chosen primarily based on proximity to airports and land availability near the Hudson River VFR corridor. In addition, the three heliports in the southern part of Manhattan were also included as they currently transport tourist, flights into airports, and flights to the Hamptons.

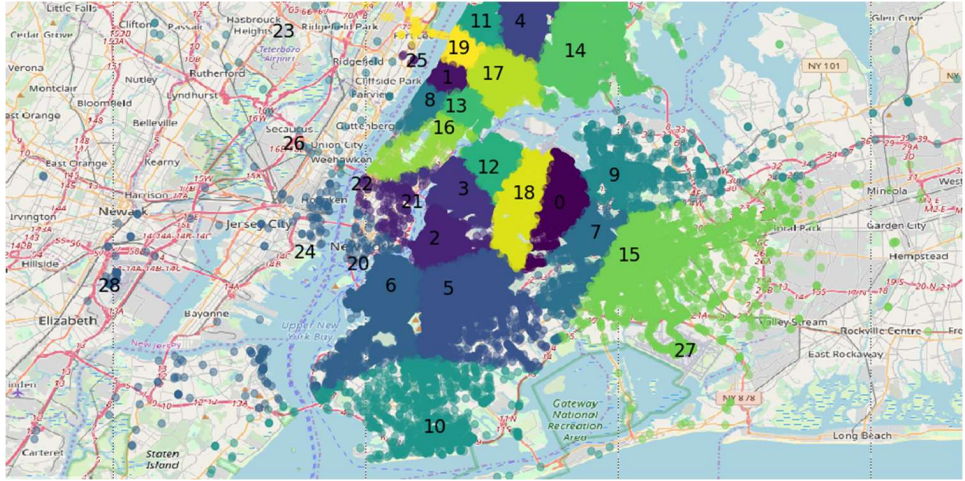

**Figure 2. Vertiport locations and their associated catchment area for a year of demand data.**

A module was developed to allow creation of custom network structures given a metropolitan region map, corridor rulesets (e.g., altitudes, spacing, over land flight objectives), and vertiport coordinates. However, to limit the scope of the paper we assume near term UAM operations would leverage the existing set of helicopter routes within the NYC airspace because early operators are expected to leverage these routes [4], [6]. Figure 3. shows the corridor network designed to utilize existing helicopter VFR routes in the NYC airspace. Red nodes represent vertiports while blue nodes represent the nodes along the flight path. After the corridor network has been created, the scheduler uses Dijkstra's algorithm [23] to find the shortest path between every set of vertiports. Using this network, the scheduler will choose the shortest route possible from an OD pair, so sectors of the network will be underutilized.

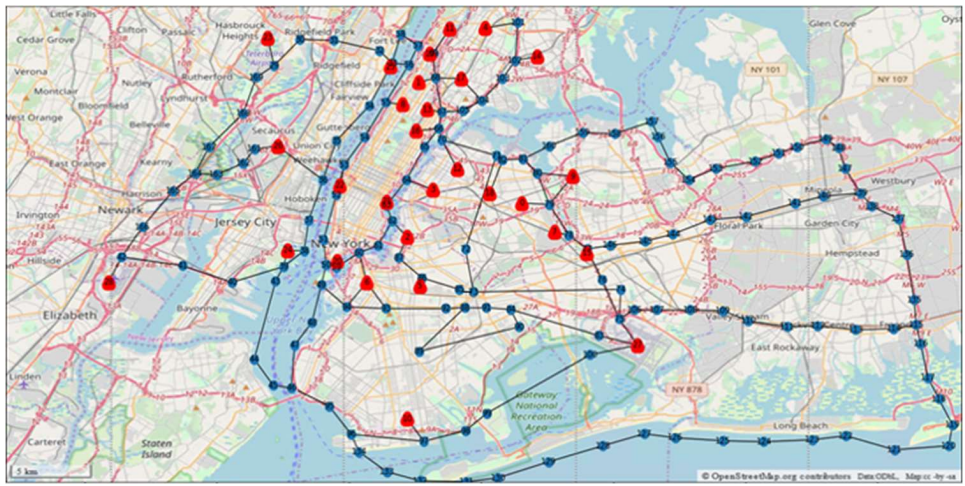

**Figure 3. Vertiport and corridor network structure designed to represent existing helicopter routes in NYC airspace.**

D. Aircraft Scheduling

The scheduling component of the model outputs a list of flights, during a set period of time, that satisfy the given demand. To accomplish this, the scheduler iterates through the demand on a per minute basis assigning demand to aircraft until either all aircraft are full or there is nothing left to assign. The output is a list of trips consisting of aircraft identification, departure time, time of arrival, departure vertiport, and arrival vertiport. Areas not addressed by this module are the number of vertipads at each vertiport, flight separation standards, and turnover rates at vertiports. However, there is an inherent one-minute time period between landing and takeoff (e.g., minimum time step).

Theoretically, every aircraft could simultaneously be assigned to fly to one vertiport, from the same vertiport, at the same time, as long as it followed the demand model. The scheduling module inputs consist of the fleet size, aircraft cruising speed distribution, aircraft-vertiport distribution, demand data, and the corridor structure. The aircraft-vertiport distribution indicates the starting vertiports for each aircraft, a uniform distribution being the default. The states for each aircraft are defined in Table 1; all aircraft start in state zero (awaiting assignment) with a total state set of five.

**Table 1 Aircraft States Sets Available in Scheduling Module.**

| Aircraft States | Descriptions |
|---|---|
| 0 | aircraft at vertiport with no demand assigned |
| 1 | aircraft assigned to pick up demand at a different vertiport but not yet flying |
| 2 | aircraft assigned to transport demand from current vertiport to another but not yet flying |
| 3 | aircraft assigned to pick up demand at different vertiport and flying |
| 4 | aircraft assigned to transport demand from current vertiport to another and flying |

Demand is considered up to the current time step with a given resolution and is not forward looking. Each vertiport has a weighting factor used when scheduling, such that priority is given to the vertiport with the highest weight: it is equal to the summation of the accumulated delay of all the passengers waiting at that vertiport. The first strategy involves assigning aircraft in state zero, one, or two, where if in state one or two the requests must have the same destination as the aircraft. Aircraft in state zero are able to reposition from a vertiport without demand at a specific time to pick up passengers in an underserved vertiport. When multiple aircraft are found in this state, the aircraft with the most open seats are selected. The second strategy prioritizes aircraft in state one or three that are scheduled to pick up demand at the vertiport of interest and that have the same destination. Aircraft in state one assigned in this way will also take any available demand from its current vertiport to the vertiport that it is assigned to pick up from, regardless of the weight of its departure vertiport. If the aircraft is in state three, it will simply pick up the demand once it has arrived. In the case where there are no aircraft able to satisfy the demand, the demand is held to be checked again later.

Parties of people are not taken into consideration when checking demand. For example, if there are four requests at the same time at a vertiport, and there are four aircraft with one open seat each, these four people will be split among the aircraft. This example only holds true if the aircraft are headed to the same destination. However, if there are four open seats in one aircraft, all four requests will be assigned to this single aircraft. Future implementations are expected to maintain parties together. An example scenario involving a system of two vertiports and a fleet of four aircraft is shown in Figure 4.. Once an aircraft is assigned, its travel time is calculated. By default, this is done by using the pre-calculated corridor distances between the vertiports, and dynamics simulating takeoff, landing, and cruising. As an example, in our 29 vertiport system, the mean corridor distances between vertiports is 18 nautical mile (NMi), with a maximum of 41.8 NMi and a minimum of 3.1 NMi. Direct flight paths between vertiports can also be used to calculate travel time and this is done within the scheduler. After assigning all possible demand for a time step, aircraft states are updated.

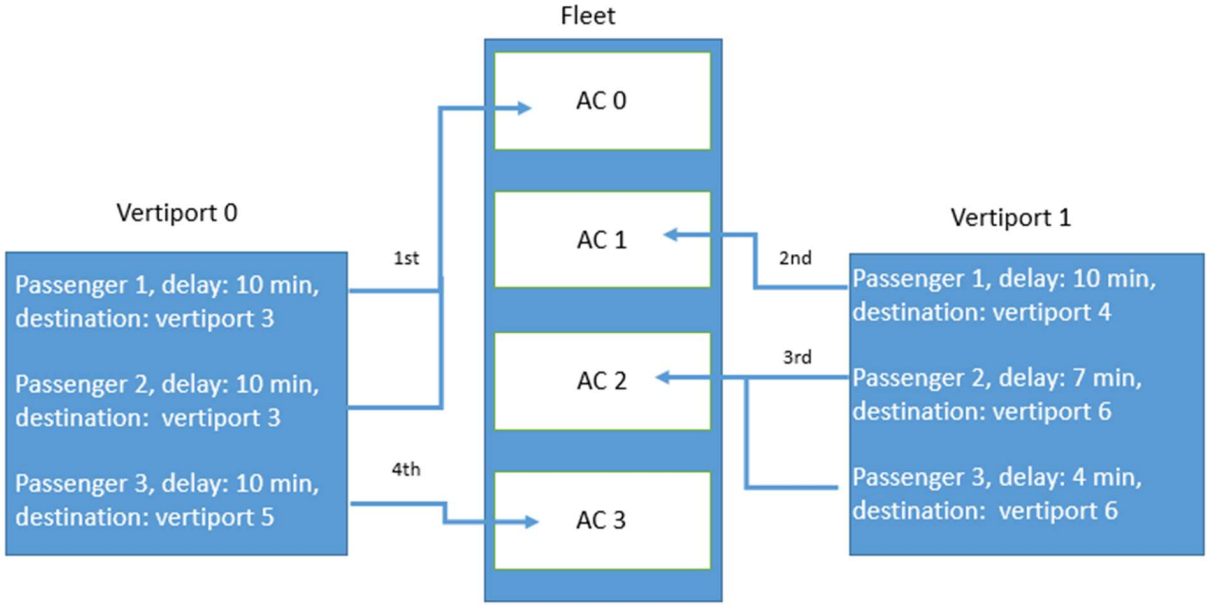

**Figure 4. Example scheduled demand assignment for from the greedy scheduler.**

### E. Discrete Event Simulation

A discrete-event simulation model was developed to analyze the system performance of UAM traffic networks. Applying a discrete-event model allows the system to jump to the next time step where a new event occurs, with the assumption that the state of system does not change outside of the event queue until the new event. The simulation views each aircraft and each vertiport as an individual resource with a fixed capacity defined by the number of seats in the vehicle and the number of vertipads in the vertiport respectively. When the demand for a resource exceeds the available capacity the excess requests for the resource are put in a queue and serviced on a first-come-first-served basis. In the case of the vertiport the excess aircraft represent the flights that are loitering near the vertiport when the vertiport has no additional parking spaces. Additional service times can be added to account for the time spent unloading and loading passengers. This simulation supports analysis of various scenarios by modeling the traffic events defined in the schedule. A scenario is defined by configuring the number of vertiports, the number of vertipads at each vertiport, and the vertiport locations. The user can define the vehicle characteristics by specifying the number of aircraft in the fleet, the cruising speed of each aircraft, and the number of seats on each aircraft. The simulation also incorporates constraints such as the holding time at the vertiport prior to releasing flight and the minimum amount of time each aircraft needs to spend on the ground before departure.

At the beginning of each simulation, aircraft are placed at their initial vertiport defined in the schedule. Flights are scheduled on-demand and travel between vertiports in order to meet the requested trip demand using the scheduler described in Section III.D. At each vertiport, passengers are loaded onto the vehicle prior to departure. The time associated with loading can be represented as a constant or by a user specified distribution. When a flight departs from a vertiport, it flies along the set of waypoints within the flight plan. At each waypoint, an event is triggered and a new decision is made regarding where to proceed. If the aircraft receives no new information, it will proceed along its current flight plan. When a flight approaches the destination vertiport, it will land if there is a vertipad available and passengers are dropped off, otherwise it will hold until one becomes available. A diagram of the simulation framework is shown in Figure 5..

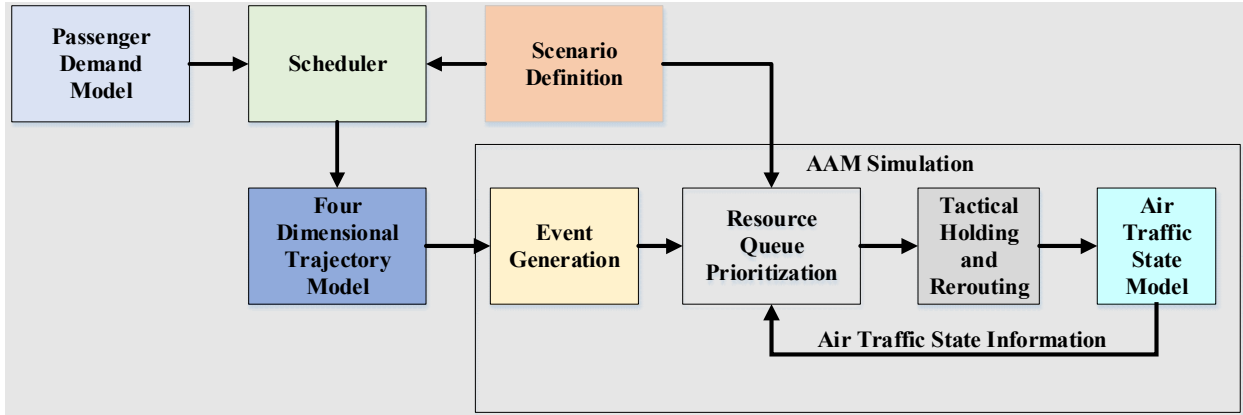

**Figure 5. Flow chart of the complete AAM Simulation Framework.**

The simulation evaluates the decisions of the users through a set of operational metrics including passenger delay, ground delay, air delay, vertiport utilization, and flight diversions. At the end of the simulation, the users can view the performance of the scenario relative to these metrics. By iteratively running different configurations against various scenarios, users can gauge the effectiveness of potential interventions and gain insight into how to tailor the schedule, fleet size, and vertiport structures to optimize the metrics of greatest importance for a given passenger demand distribution. They can also perform system trade studies to understand the relative variation between scenario parameters. Ultimately, network designers and regulators can use the results of these studies to configure the AAM network to meet specific performance requirements.

## IV. Analysis

A set of analyses are conducted to evaluate the performance of modeling framework described in the previous section. With this corridor network and set of vertiports locations, we establish experimental parameters that are explored in a parametric sweep. We first explore the maximum number of aircraft that would occupy the airspace given an assumed fractional replacement of ground vehicle demand. This is accomplished by sweeping the fleet size available for each vehicle capacity with no resource constraints for vertiport landing areas or turnaround times. Second, applying resource constraints to the vertiport landing areas, we explore the impact of vehicle passenger capacity on the number of vehicles simultaneously in air and its impacts on trip delay, number simultaneous operations, vertiport throughput, and the required fleet size to compete against the total travel times in the taxi database. Last, the interactions this network may have with surrounding airspace traffic are explored. In total, the Monte Carlo approach requires a total of 5.28 million simulations from the combination of parameters in Table 2. Note, the cruising speed of aircraft was randomly drawn from a uniform distribution between 87 and 120 knots as a review of available open source information denoted this as an applicable range for operations at low altitude [3]. [4], [5], [28]. While the framework is implementable on a single personal computer desktop, to achieve this large parameter sweep, the framework's modular structure was exploited to implement it within the MIT Lincoln Laboratory Super Computing Center (LLSC) [24]. By leveraging the LLSC, complex model interactions can be simulated and detailed resource constrained conclusions can be determined that previous studies could not.

**Table 2. Variables explored in analysis and their associated values or range.**

| Parameter | Value |
|---|---|
| Percent demand replacement | 5% - 100% |
| Fleet size | 0 - 6,000 |
| Vehicle passenger capacity | 4, 6, 8, 10 |
| Extra parking spaces at vertiports | 0 - 11 |

### A. Unconstrained resource analysis

In this analysis, a full day of demand is queried for a percentage of replacement of taxi traffic, with the maximum 100% replacement accounting for 2,139,005 passengers transported for the day. An example of the variation in the demand throughout the day for is shown in Figure 6.. The demand example shows a high number of requests at the expected peak times of morning and afternoon commutes with a drop off in requests late in the night. This demand is then used by several scheduling instances each with a set fleet size and vehicle capacity assumption which produces a days' worth of operations that are then simulated without resource constraints. Note this initial assessment provides a minimum requirement as the vertipad availability resource is unconstrained. As seen in Figure 7., demand replacement percentage directly controls the number of simultaneous aircraft in the air required to meet the demand. Under the common four passenger vehicle configuration that industry is converging to for their business case, the analysis shows a need for fleet size on the order of ~200 to ~6000 aircraft. The greedy scheduling approach complements a passenger pooling strategy as higher capacity vehicles reduces the fleet size required to ~150 to ~4000 aircraft. On average, an increase in 2 passengers per vehicle leads to at least 7% reduction in aircraft simultaneously in air. With the initial increase from four passenger fleets to six passenger fleets having the largest reduction in simultaneous aircraft operation of 50%. Due to the unconstrained nature of this analysis, it cannot be concluded that replacing 5% of the ground for-hire vehicle demand with UAM would result in 150 - 230 simultaneous vehicles in-air. In order to accomplish this task, a sweep of resource constraints is required. However, it is evident even a 5% replacement with unconstrained resources would stress the New York airspace. Using the approach derived by [24] shows the 60 NMi Mode-C veil around KJFK International airport sees 60 –800 aircraft operating at once depending on the time of day. Even if this airspace density was sustainable, the AAM vertiport network would see high demand vertiports experiencing throughputs on average of 562 landings and takeoff per hour with peak times seeing 1,099 landings and takeoff per hour. When looking at the throughput that must be sustained by each vertiport, analysis indicates operations at a single vertiport exceed the combined number of landings and takeoff of the four major airports in the vicinity (i.e., JFK International, Newark, La Guardia, Teterboro) which experience an average of 80 landings and takeoffs per hour [26].

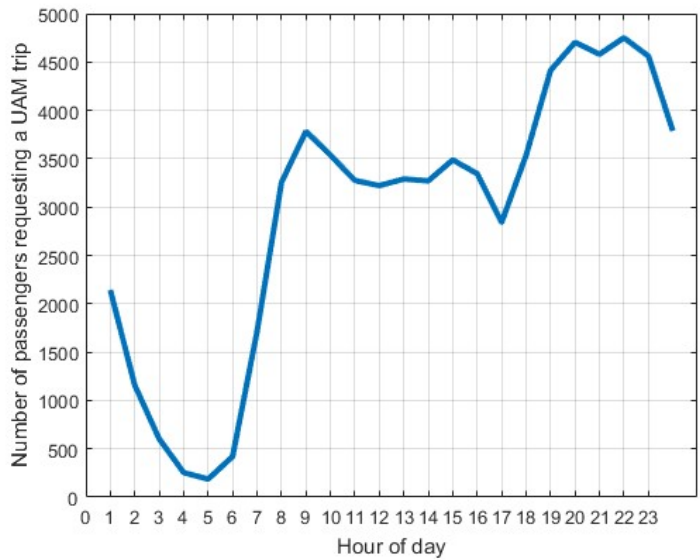

**Figure 6. Temporal variation in the number of requests for UAM modeled as a direct replacement of 5% of the ground taxi.**

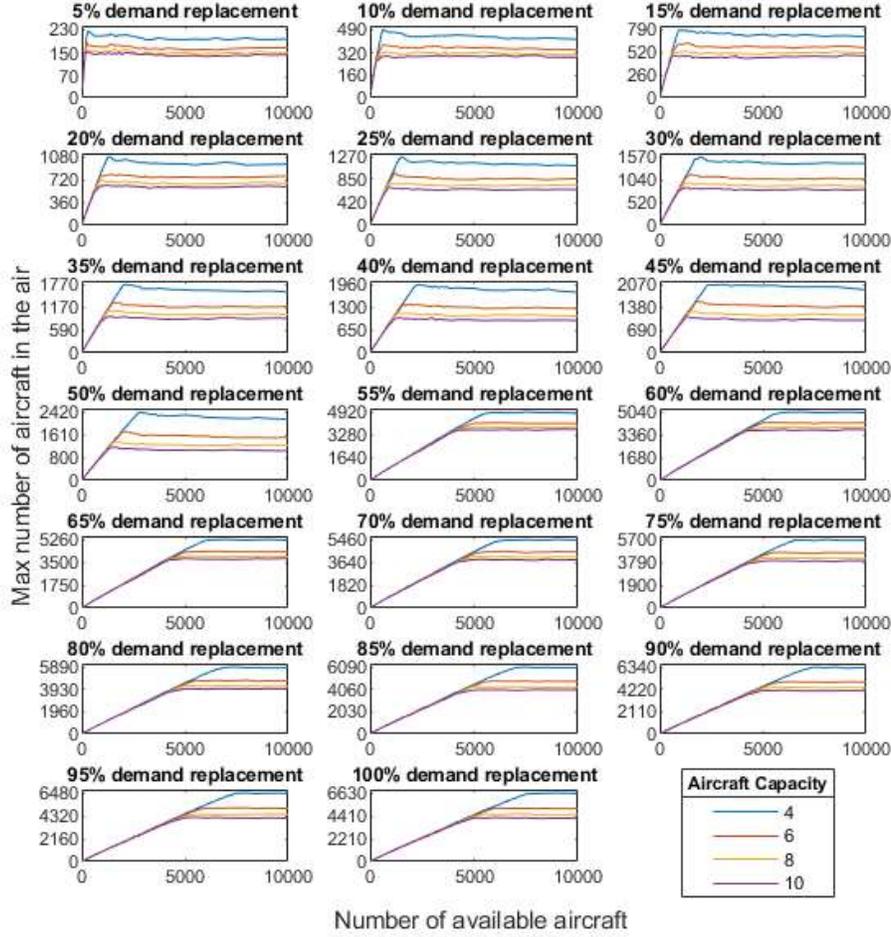

**Figure 7. Simultaneous number of aircraft in the air as a function of demand replaced with UAM and fleet size.**

**B. Constrained resource and delay analysis**

The impacts of resource availability are explored by constraining the number of parking spaces at each vertiport as expressed by equation 1 and increasing the number of extra vertipads at each vertiport. The authors recognize that at the extreme case a fleet size of 6,000 aircraft would require all vertiports to contain 217 vertipads. It is unlikely that major vertiports could sustain 217 vertipads in the near future, however the scenarios are included to understand the relationships in the parametric sweep.

$$Vertipads\ per\ vertiport = \left\lceil \frac{Fleet\ Size}{Number\ of\ Vertiports} \right\rceil + Extra\ Vertipads \quad (1)$$

There remains an uncertainty on whether UAM operations will swap batteries or develop quick charge technologies that could reduce turn-around times to less than 20 minutes [27]. It is expected this constraint will approach similar turn-around times of helicopter operations on the order of 5 minutes or less [15]. Thus, for the purpose of this analysis the impact of turn-around time is not explored and is maintained as a constant of 1 minute. To understand the benefits of increasing vertipads at vertiports or increased passenger capacity to levels across and beyond those explored by [7], [8], [9], [10], and [11] we create schedules varying passenger capacity, fleet size, and available vertipads at vertiports while holding the demand served equal. The results show increasing parking spaces at vertipads to accommodate higher vertiport throughput has a lower impact in reducing trip delays as opposed to pooling passengers into larger vehicles, Figure 8.. For example, with zero extra vertipads available, a fleet size of 1,500 aircraft composed of four passenger capacity vehicles would experiences a trip delay of 29.7 minutes. However, increasing the passenger capacity to six or eight passengers reduces it to 10.9, and 3.6 minutes respectively (i.e., 63% to 88% reduction in average trip delay). Note, further increasing capacity past six passengers shows diminishing returns as the fleet wide passenger capacity outpaces the demand for a trip. Operating the 1,500 aircraft fleet with 11 vertipads resulted in an average trip time of 14.64 minutes. Therefore, to achieve an equivalent reduction in delay similar to increasing passenger capacity from four to six passengers requires more than 11 extra vertipads per vertiport. Therefore, unlike the unconstrained case (Section IV.A) where a fleet size less than 500 aircraft would meet the demand, when applying resource constraints, a larger aircraft fleet is required for a greedy scheduling approach to reduce the average passenger and trip delay to values below 15 minutes. The data also shows the inefficiency of the greedy scheduler as it can results in suboptimal use of the aircraft and result in larger delays

In particular we find a fleet size of 1716, with no extra vertipads, was the first resource constrained simulation to reduce the average trip delay and the average per passenger delay below 15 minutes, Figure 9.. We concentrate on this fleet size as further delay would be expected to impact passenger willingness to travel [28], [29].

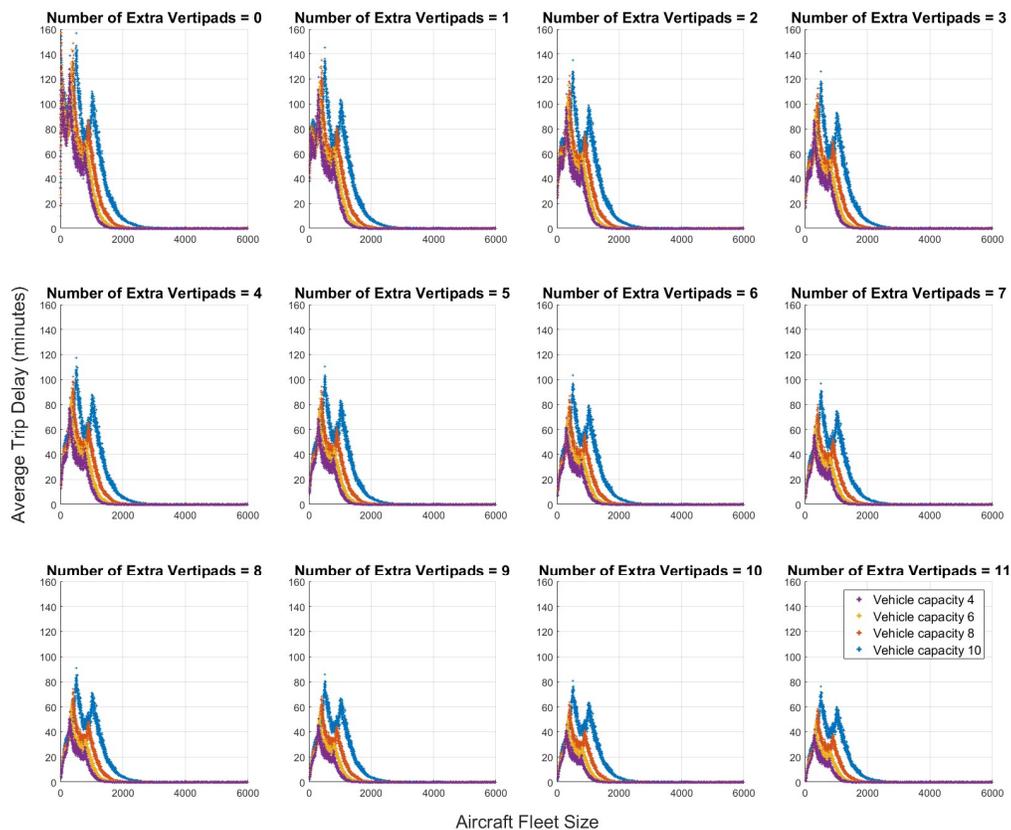

**Figure 8. Delay experience by a scheduled trip given a fleet size and allocating extra vertipads per vertiport.**

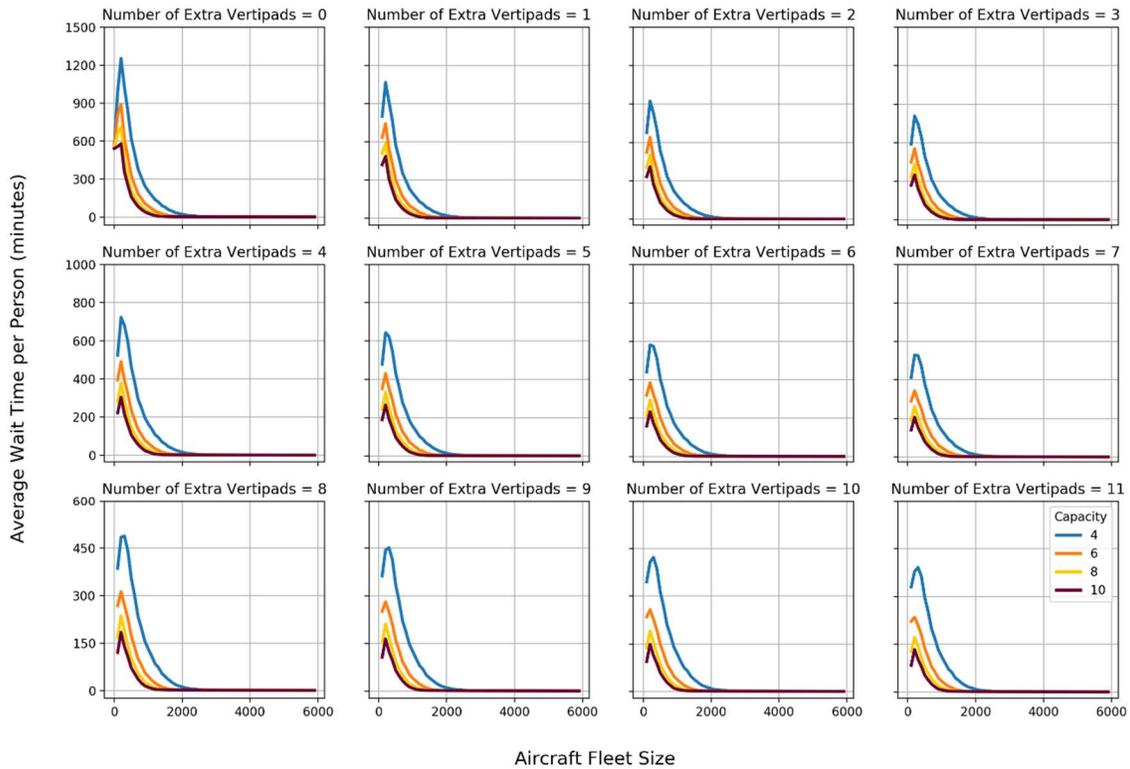

**Figure 9. Average delay per passenger transported for a given fleet size and impacts of allocating extra vertipad per vertiport.**

An approximation of the average trip time passengers experience must account for the 1) the average time it takes passengers to arrive at the vertiports, 2) onboarding and offboarding delay, 3) UAM trip delay to reach destination vertiport, 4) the average time it takes to arrive their destination. The average travel time to access and egress a vertiport depends on the transportation mode used for this portion of the trip, as well as the location of said vertiport. In this paper, we calculate an average time to arrive and depart based on the average trip time required for ground vehicles for-hire [21] operating within the catchment area of the vertiport at the peak traffic time of 8am. Thus, 1) can be approximated as the average ground transportation time of 13 minutes with a maximum of 240 minutes, and minimum of 5 minutes to travel from the true origin and destination of a passenger to each respective vertiport, see Figure 10. for details of each vertiport. To constrain the scope of this study we hold the onboarding and offboarding time constant for simulation at 5 minutes. On average a multi-modal trip would be composed of two 13 minute taxi trips, with a average 20 – 30 minute UAM trip, which itself can have on average 15 minutes of delay for a total of 46 – 56 minutes. As a comparison to ground transportation, analysis of the NYC database [21] shows the average trip time for a taxi passenger is on average 20 minutes in length with the maximum trip delays seen in a week upwards of 150 minutes. The authors recognize a demand capacity balancing scheme could reduce the delays experienced by single passenger, and these results communicate a need for optimization in scheduling for UAM operations as resource availability can create unattractive total trip times compared to ground transportation.

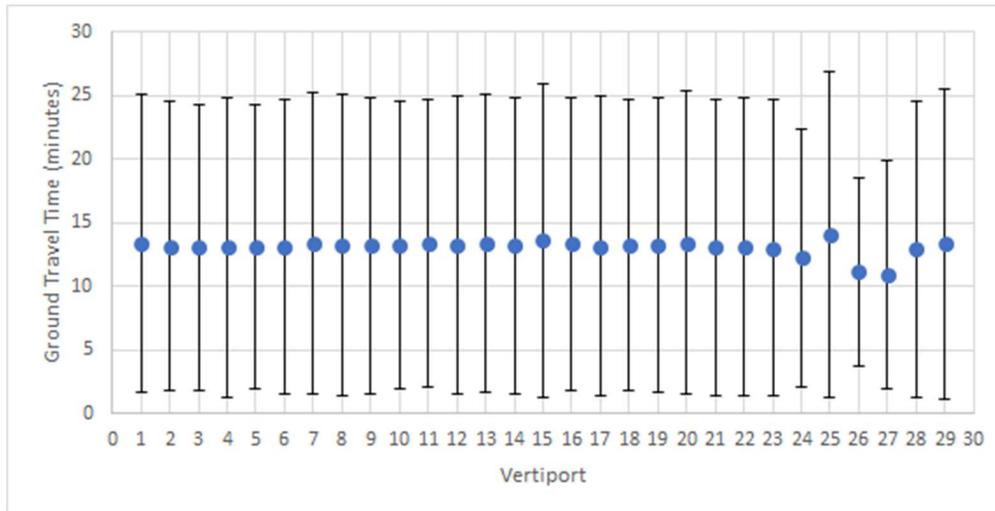

**Figure 10. Average ground taxi travel times to and from the respective vertiports and its catchment area.**

The load factors, empty flights, and total simultaneous operations in air of the 1683 fleet size are explored to understand the temporal relationship of these metrics as well as their evolution across the day. As seen in Figure 11. The total number aircraft in air per hour, the number of empty flights per hour, and the fleet wide load factor per hour for a resource constrained simulation with a fleet size of 1,683 and fleet composed of four passenger aircraft., the load factor drops below 0.40 between 3 – 5AM since the greedy scheduler chooses to assign ~100 aircraft to meet the low demand in this time frame. This load factor is predominantly driven by the spatial sparsity in the demand, as seen by the percentage of empty flights compared to all flights, and the fact our fleet size remains equal through the day. That is, vehicles are not taken out of service and passengers are not pooled to increase the load factors, as might be the case in a revenue maximizing scenario. As demand increases with the morning commutes a smaller percentage of flights are conducting aircraft repositioning operations, leading to an increased utilization of the vehicle fleet and passenger pooling. As seen in Figure 11, the simultaneous number of aircraft in the air never reaches the total fleet size available. Unlike the unconstrained analysis, the number of aircraft required to meet a direct replacement of 5% of for-hire vehicles would increase the total number currently operating in the airspace by 2 – 3 times.

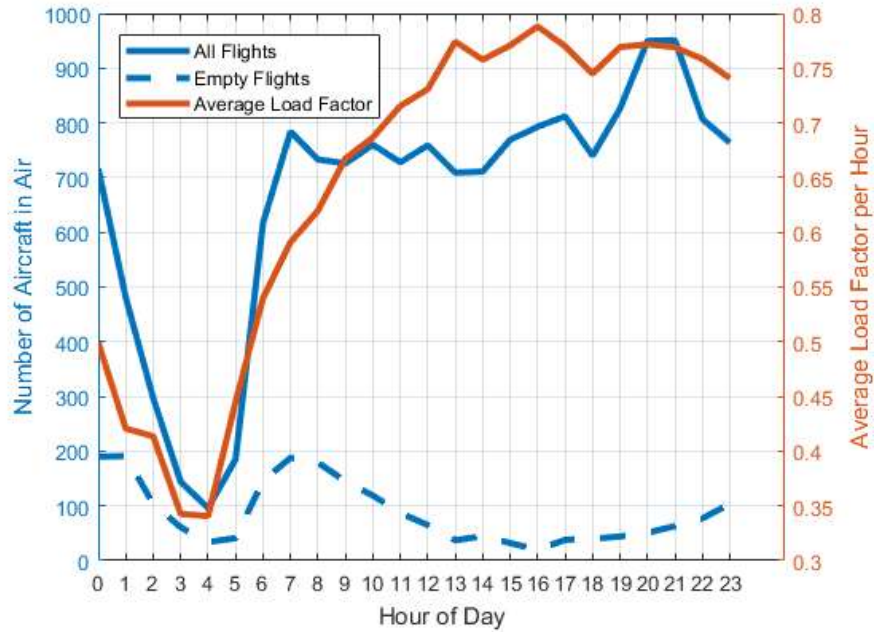

**Figure 11. The total number aircraft in air per hour, the number of empty flights per hour, and the fleet wide load factor per hour for a resource constrained simulation with a fleet size of 1,683 and fleet composed of four passenger aircraft.**

### C. Trip Delay Patterns

The previous sections examined the effect of mismatches between demand and capacity by averaging delay over the entire network. This data shows the aggregate impact of these demand/capacity imbalances on network operations but does not describe the variability of the impact of the locations where the delay burden is greatest. Traffic demand patterns are not uniform across urban areas and accordingly it is important to understand which trips and vertiports might be most impacted when the system does not have the capacity to accommodate the demand. We conducted an additional parametric sweep in order to gauge the effect of aircraft capacity on vertiport delay. Using the results from section IV.B. we selected the scenario with 4 seats per aircraft that yielded the maximum average delay such that the average delay was less than 15 minutes while adding no extra vertipads at any vertiport. The scenario has a fleet size of 1716 vehicles and used 29 vertiports. The number of extra vertipads at each vertiport was incremented from 0 to 10. The resulting average delay between each vertiport pair for two of these cases is shown in Figures 13. For the purposes of comparison, another scenario with 6 seat per aircraft and the same number of vehicles was also performed. The resulting performance for two cases is shown in Figure 14. The figure indicates whether the average delay at a specific vertiport pair is greater than 15 minutes using two markers. The purple markers indicate that the delay exceeds 15 minutes while the white markers indicate that average delay is below 15 minutes. In some cases, there was insufficient demand a certain origin vertiports to schedule any trips. In these instances, the relationship between origin-destination pair delay is not shown on the figure.

The figure suggests that the delay impact of resulting from insufficient capacity for the assumed demand is widespread but not uniform across the city. The delay is particularly prevalent at vertiports 20, 21 and 22, which are located in Lower Manhattan where the financial district is located. This is not surprising as the trip demand is proportionally greater relative to the rest of the city. This delay is present throughout all of the 4 seat aircraft scenarios regardless of whether there are 0 or 10 excess parking spots. When the number of seats in the aircraft increases from 4 to 6 there is a substantial drop in the number of origin-destination pairs that have an average trip delay exceeding 15 minutes. Although the Lower Manhattan vertiport pairs are still largely affected, there are significant reductions at vertiports 0, 1, 2, 3 and 4 located in parts of Brooklyn and Upper Manhattan as well as vertiports 27 and 28 where Newark and John F. Kennedy International Airports are located. This general trend is present whether there are 0 or 10 excess parking spots, though there is some improvement at nearly all vertiports will fewer trip pairs associated with 15 minute or greater delays. The results suggest that increasing the passenger capacity of the air taxis may provide greater benefit for the delay performance within the network than increasing the number of parking spots at each vertiport. It should be acknowledged that demand could also be controlled by varying the price of a trip. Given the excess demand seen in the taxi data for trips to and from Lower Manhattan we would expect the prices for these trips to be significantly higher. It is quite conceivable that such a price increase could limit demand to a more tolerable level reducing the overall delay. We leave it to future researchers to study the effects of these modifications.

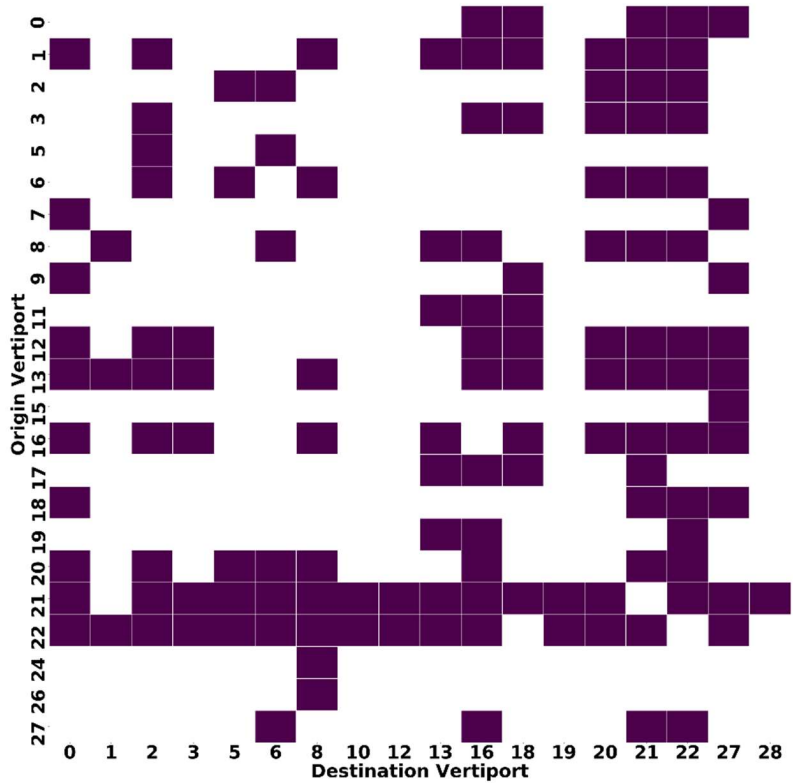

**Figure 13a. Vertiport pairs with an average delay greater than 15 minutes with 4 seats per aircraft scenario and no additional vertipads per vertiports.**

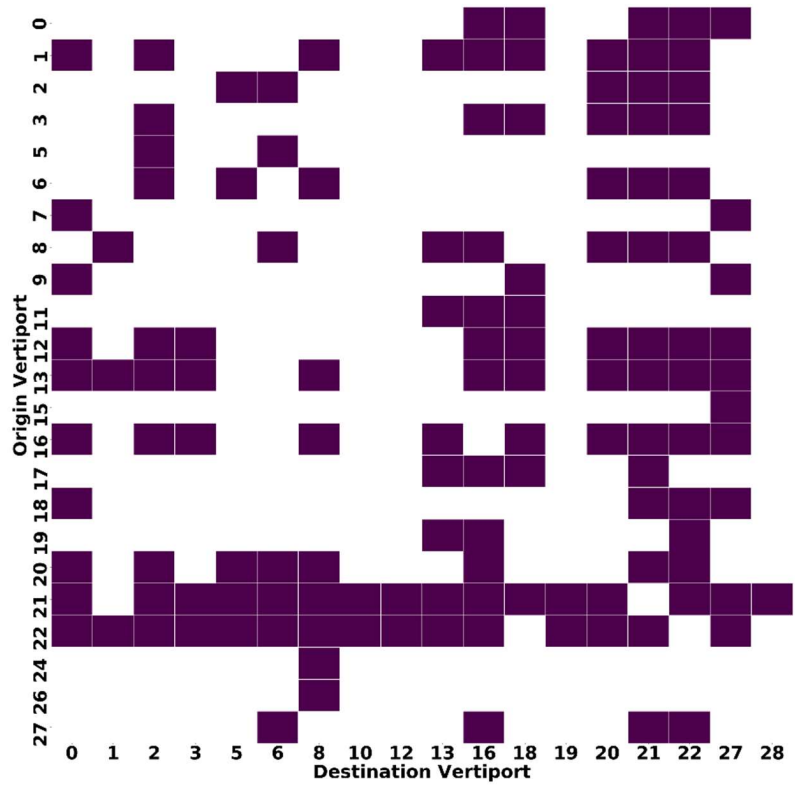

**Figure 13b. Vertiport pairs with an average delay greater than 15 minutes with 4 seats per aircraft scenario and 10 additional vertipads per vertiports.**

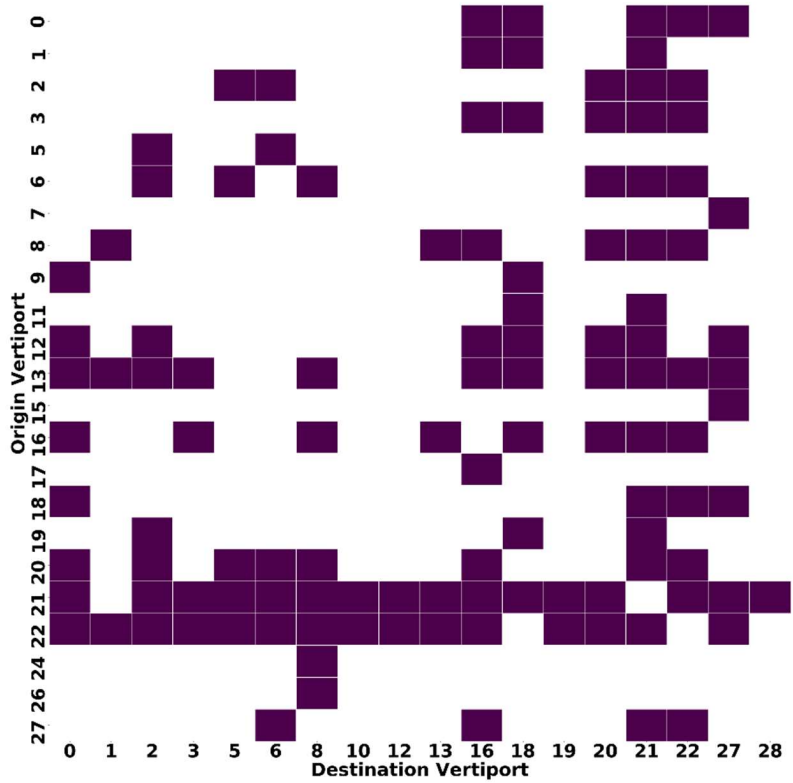

**Figure 14a. Vertiport pairs with an average delay greater than 15 minutes with 6 seats per aircraft scenario and no additional vertipads per vertiports.**

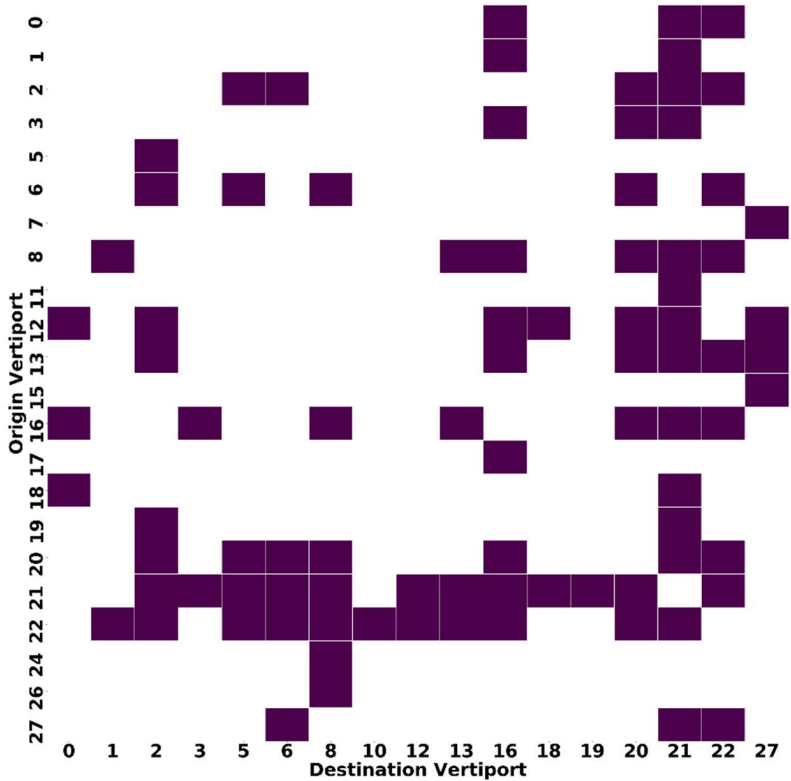

**Figure 14b. Vertiport pairs with an average delay greater than 15 minutes with 6 seats per aircraft scenario and 4 additional vertipads per vertiports.**

### D. Network sector utilization and considerations for designing UAM corridors

Prior sections have focused on the aircraft, and airspace utilization required to service the underlying demand and understanding systems characteristics that cause UAM to be less competitive as a transport. In this section we focus on the spatial density across varying sectors of the network. Its understood by the authors that the actual density values are representative of the demand being met; however, in our lowest demand replacement we see a high-density region due to 1) the demand at nearby vertiport, 2) common network sectors used by transiting traffic. As described in Section.III.D the simulation framework assigns the shortest route between vertiport pairs, thus creating common sectors for transit traffic. The common transit sectors combined with the high demand for wall-street vertiports traffic causes the hotspot in the lower region of the Manhattan, see **Error! Reference source not found.** and Figure 12.

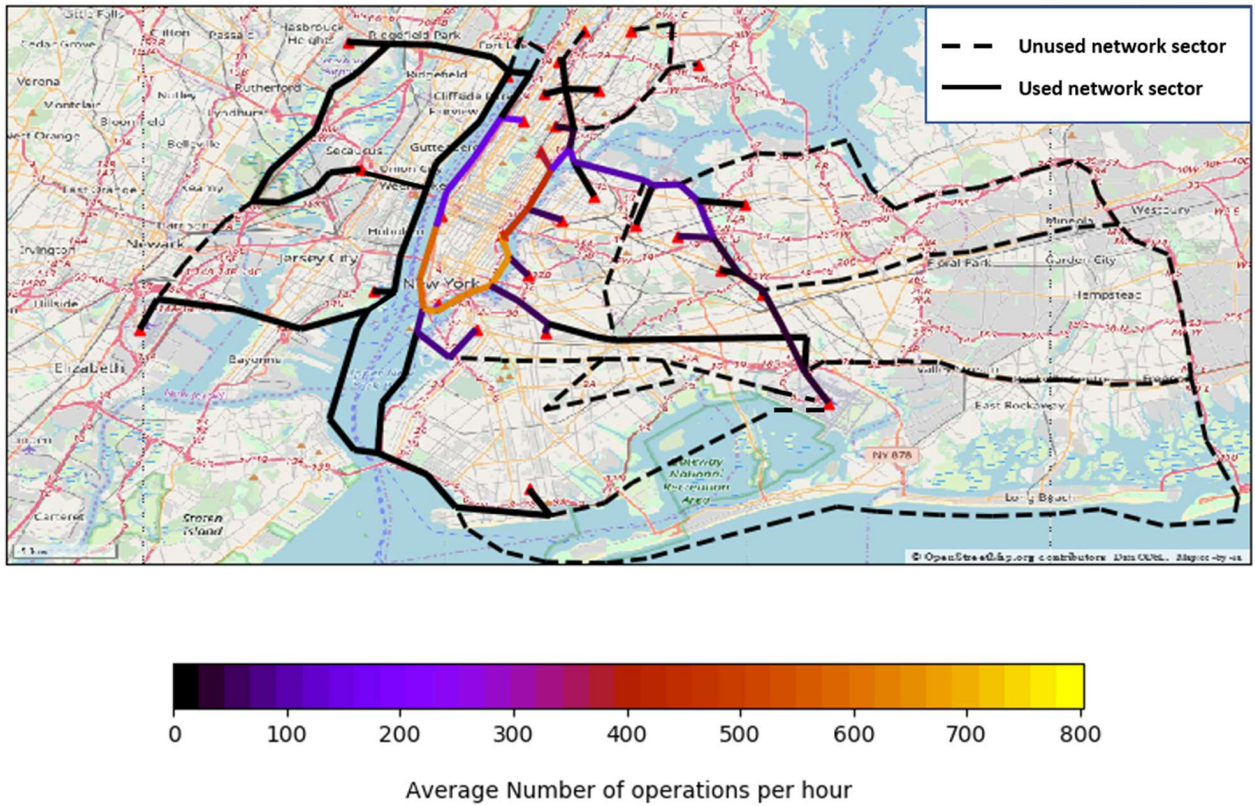

**Figure 12. Average traffic density per hour accross the network.**

The design of our UAM network, particularly the location of the vertiports, was independent of any considerations for current manned aircraft operations in the New York City airspace. As noted in the paper, the authors sought to model and optimize a UAM corridor network to satisfy underlying demand. Given this network, we started to assess interactions with aircraft in the surrounding airspace and identify potential challenges. A full analysis of how a specific, not notional, UAM corridor network integrates into a specific section of the NAS is identified as future work and the focus of a future paper.

To support this analysis, we identified three representative aircraft: a rotorcraft operated by a local tour operator (Bell 206L-3), a commercially operated fixed-wing single engine seaplane (Cessna 208b), and a fixed-wing multi-engine operated by a major American airline (A321-231). Observations of these aircraft were sourced from the OpenSky Network [30], a community network of ground-based sensors that observe aircraft equipped with ADS-B, and processed according to the workflow described in [31]. Processing steps consisted of (1) parsing and organizing the raw data; (2) archiving organized data; (3) processing and interpolating into track segments. Processing software can be found in the open-source repository [32].

The curated dataset and processed track segments were the same used to train a correlated model of aircraft operating out of aerodromes in the terminal environment. Temporally, the dataset consisted of the first 14 days of each month from January 2019 through February 2020. The scope included, but was not limited to, New York City and the surround airspace. Refer to [33] for more details where this dataset is referred to as the "aerodromes dataset." Figure 13 illustrates the processed tracks near the proposed vertiports and Figure 14 provides a more detailed map of the rotorcraft. These figures illustrate approximately 87.9, 2.4, 4.7 flight hours respectively for the rotorcraft, seaplane, and larger fixed-wing multi-engine. Specifically, this analysis only considered the airspace at or below 5,000 feet AGL and within a bounding box with latitudes of 40.54981 and 40.8751 and longitudes of -73.5363 and -74.26003. As these tracks are based on only one surveillance source and collected via crowdsourcing, we do not assume the aircraft were comprehensively observed; rather these tracks are representative of the operations and behavioral trends of the specific aircraft. Notably, Figure 13 illustrates how differently each aircraft operates in the NAS, with Figure 14 showing that the rotorcraft primarily operates around downtown Manhattan while utilizing the airspace above both rivers. The behavioral trends align with the author's expectations. The rotorcraft operated by a tour company operated near popular tourists features of interest; the commercially operated seaplane predominately operated from a seaport (6N7) in the river and conducted relatively short transport operations; and the commercial carrier aircraft operated out of the large Class B airport while minimizing co-altitude or low altitude interactions with tall structures and obstacles. There is also distinct structure to these operations, with each aircraft utilizing the airspace differently. This aligns with conclusions from [34].

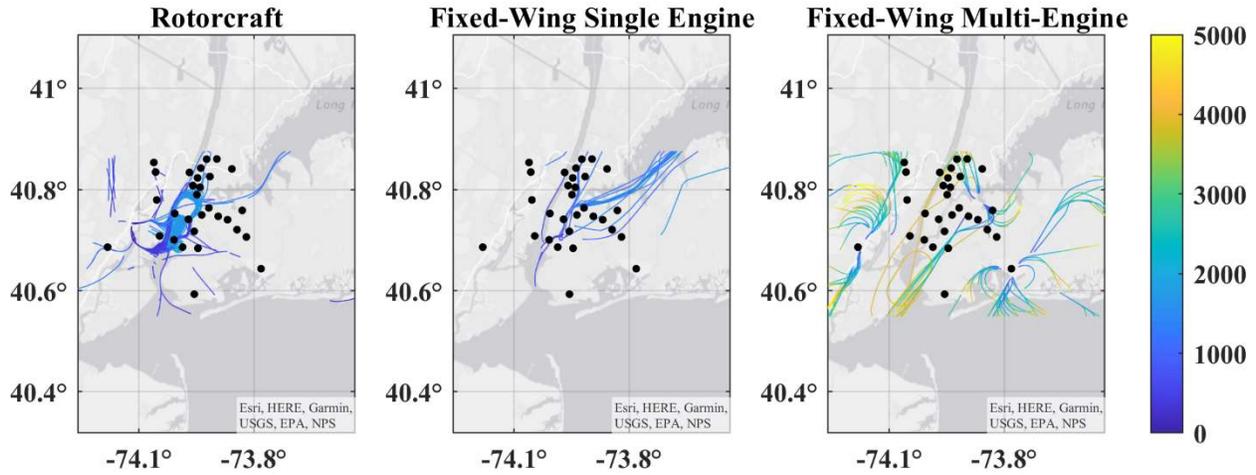

**Figure 13. Tracks of representative manned aircraft with proposed vertiport locations overlaid. Color corresponds to the estimated above ground level altitude based on ADS-B geometric altitude.**

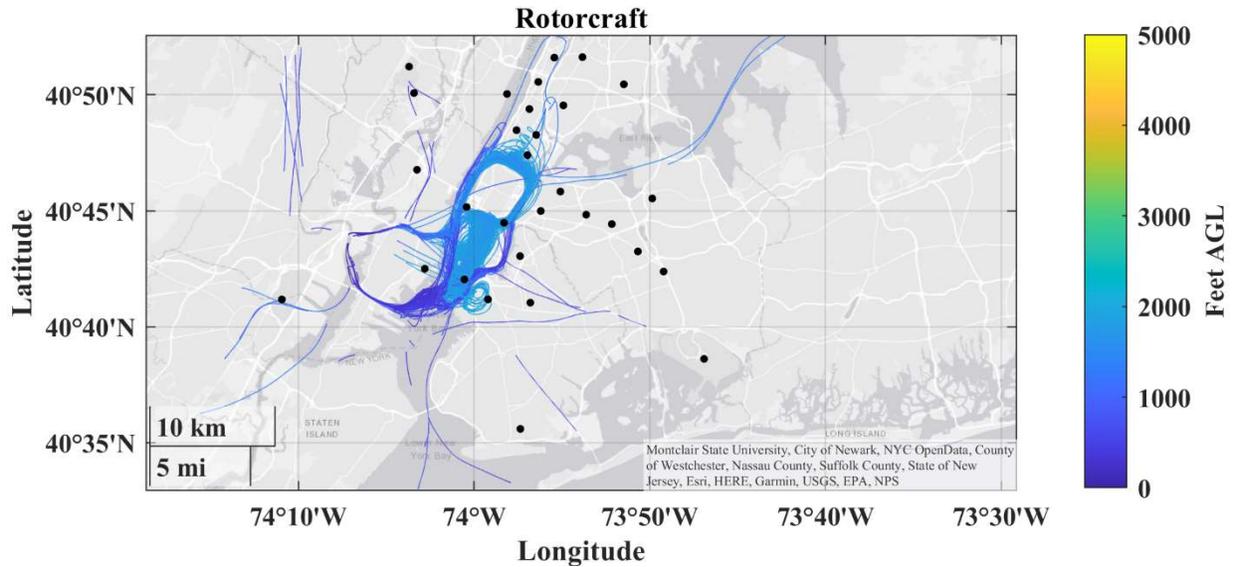

**Figure 14. Example observed activity for one helicopter with proposed vertiport locations overlaid.**

Upon visual inspection, the mapped behaviors align with many assumptions of the FAA UAM CONOPS discussed in Section II.A. The rotorcraft operates near many vertiports and at lower altitudes where UAM aircraft are expected to operate. The rotorcraft tracks also align with the proposed UAM corridors illustrated in Figure 3. The fixed-wing seaplane transitions in and out of the Manhattan seaport and may cross a few UAM corridors, while the larger fixed-wing multi-engine predominately flies at higher altitudes above the UAM corridors. However, we identified a couple challenges of integrating UAM corridors in this airspace.

First, is the potential imposing of additional requirements on existing commercial operators in complex environments, namely the ADS-B equipped seaplane. The FAA UAM CONOPS assumes that aircraft transitioning through a UAM corridor must satisfy its operational and participatory requirements, such as interaction with a PSU. Unless a PSU has ADS-B surveillance capabilities, the seaplane may have to equip another capability to provide information and meet the participatory requirements. Additional guidance is also required if the seaplane shall stop squawking ADS-B out when entering a UAM corridor and then resume squawking upon exiting. This would likely impose additional financial and operational burdens on an established commercial operation. Furthermore, if the seaplane is not receiving surveillance services from a PSU, it would wholly be reliant on detecting UAM aircraft via visual acquisition during the relatively more dangerous phase of flights of takeoff and landings. This challenge is exasperated by the surrounding environment and obstacles. The tall structures in Manhattan and Queens along the East River and the Class B surface of LGA to the northeast limits where a corridor can be positioned laterally along the East River. As the seaplane must operate to and from the river, the approach and arrival procedures will likely intersect or operate extremely close to any UAM corridor along the East River. While UAM corridors could be positioned above the approach and arrival of the seaplane, increasing the altitude of the UAM operation will also increase cost and time per flight.

Second, UAM aircraft and existing manned operations may want to operate in the same region, resulting in traffic density and flow challenges. In response, we calculated the lateral distance between the representative tracks and each of the 29 vertiports. This analysis was limited to vertiports to explore if encounters between aircraft are another potential source of trip delay. The empirical CDF of the calculated distances are provided in Figure 15. Foremost the distance distributions were aircraft dependent, with the rotorcraft typically flying closer to multiple vertiports. For all vertiports, the rotorcraft operated at least 2.5 nautical miles away for at least 50% of the time while both fixed-wing aircraft operated at least 5 nautical miles majority of the time. Conversely, the rotorcraft was always within 14 nautical miles of all vertiports, whereas the fixed-wing aircraft often operated farther away. This suggests that the traffic density and type of aircraft operating near nearby airport vertiports and UAM corridors will vary. While our UAM vertiport locations were optimized based on demand, they also are potentially inherently positioned away from larger fixed-wing multi-engine traffic but located near rotorcraft operations. This is exhibited by comparing the distance distributions for two vertiports in Figure 16.

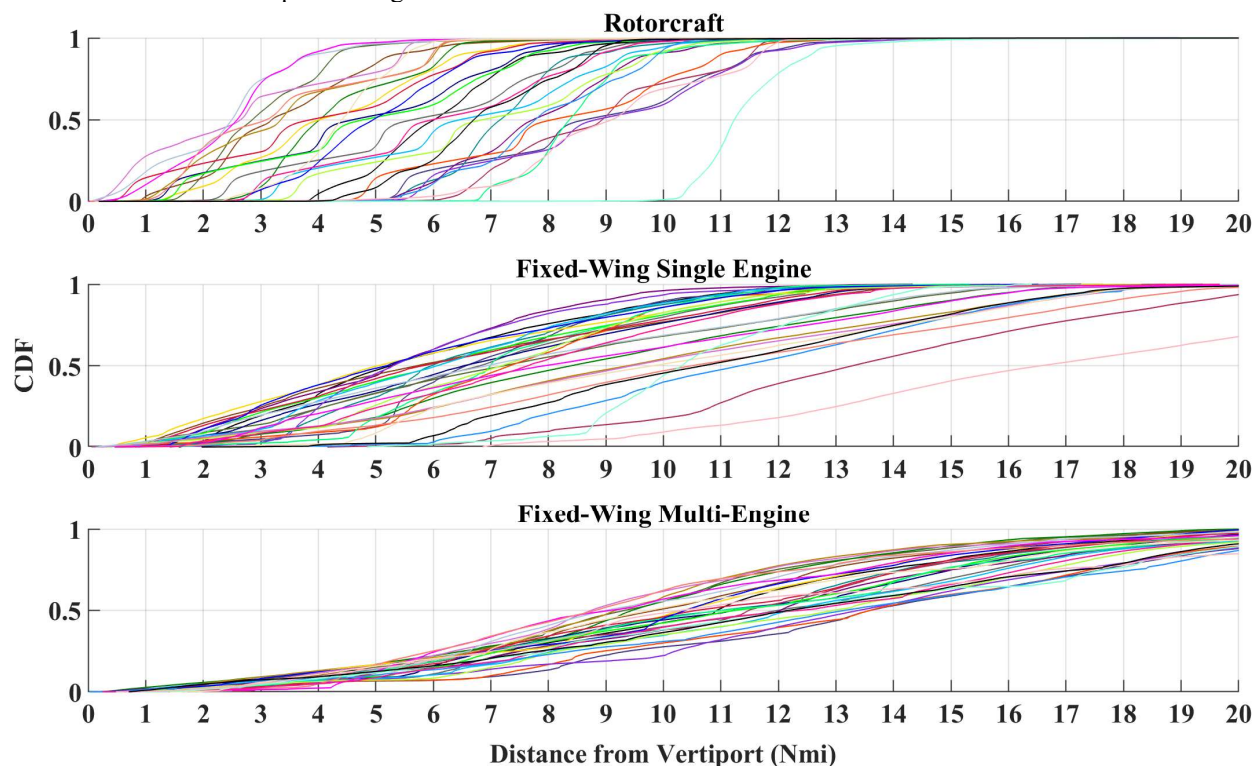

**Figure 15. CDF distributions of the distance the aircraft are from proposed vertiports. Each line corresponds to a different vertiport.**

Figure 16 compares the distance distributions for the three representative aircraft at vertiport 20 located at the Downtown Manhattan Heliport (JRB) and vertiport 28 located at Newark Liberty International Airport (EWR). While the fixed-wing multi-engine operates in and out from EWR, the air carrier use case has the aircraft quickly leave the surrounding area. This is in contrast to the rotorcraft that routinely operates close to where it takeoffs and lands. Additionally, the rotorcraft operated within 2 nautical miles of the JRB about 40% of the time. The average airspeed of the rotorcraft was 60 knots, suggesting that the rotorcraft was often 2 minutes or less away from JRB. The encounter rate between the rotorcraft and UAM aircraft could be relatively frequent, which could influence UAM corridor requirements. Characterizing potential encounters require vertical and horizontal information, so we calculated the joint CDF distribution of distance from a vertiport and the altitude of the rotorcraft.

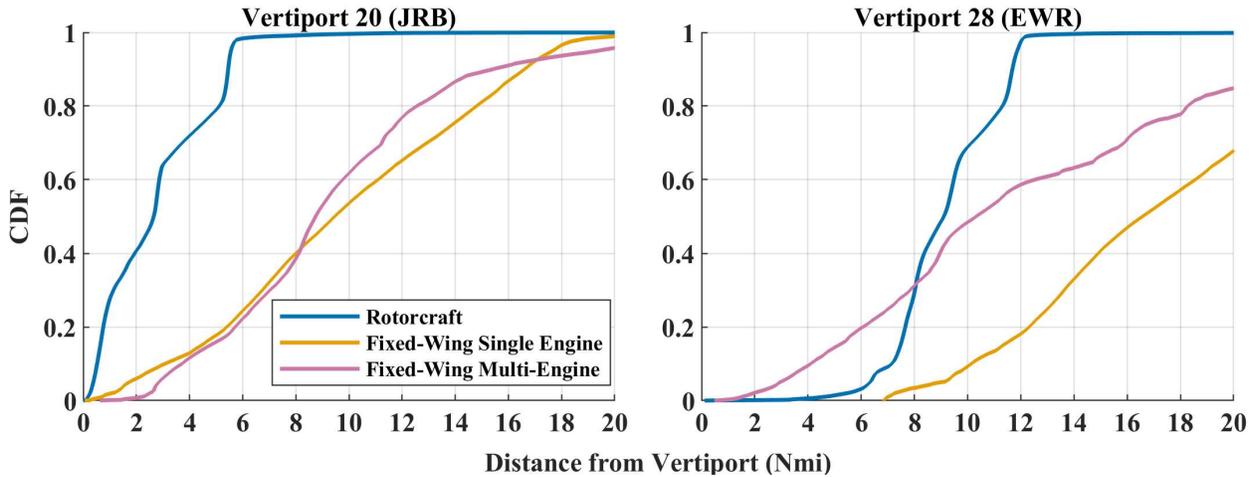

**Figure 16. CDF distributions of the distance the aircraft are from proposed JRB and EWR vertiports.**

Figures 17-19 illustrate these distributions with respect to vertiports 20 (JRB), 21 (6N5), and 28 (EWR). Notably JRB and 6N5 are only approximately 3.5 nautical miles apart but the distributions have important differences. This rotorcraft often flew lower to the ground and closer to JRB than 6N5, as illustrated by the 5% and 10% contours closer to the left and bottom in Figure 17 and farther right along the distance axis and up the altitude axis in Figure 18. So, while these two vertiports are spatially close, the potential encounter rate between this rotorcraft and UAM aircraft is much greater near JRB than 6N5. When considering UAM corridor design and requirements, could JRB and 6N5 vertiports have different requirements due to different potential encounter rates or density of other aircraft? Addressing this question is out of scope for this paper and considered for future work.

There is also an important similarity between JRB and 6N5. Both distributions indicate that 25% or more of the time that the rotorcraft is operating about 1500 feet AGL or higher. Additionally, there is typically less than a 500 feet difference between the 25th and 75th percentile contours. This indicates that the particular rotorcraft has limited variation in its cruise altitude above 1500 feet AGL. For UAM corridor design, this suggests that corridors near JRB or 6N5 are below 1500 feet AGL, that the rotorcraft would often just vertically cross the corridor. Whereas if the corridors were from 1500-2000 feet AGL, this rotorcraft would more likely operate within a corridor. If ATM wanted to minimize the potential off-nominal encounters between UAM and rotorcraft, designing corridors with altitudes below 1500 feet AGL would be a good starting point. Corridors may also be designed above the rotorcraft cruise altitude but considerations for encounters with larger aircraft would also have to be assessed.

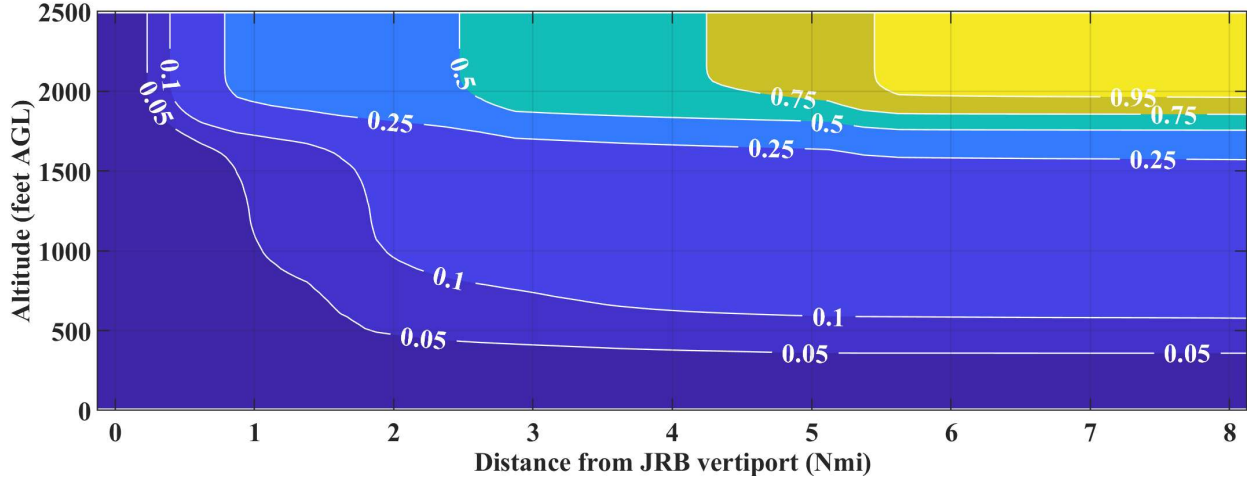

**Figure 17. Rotorcraft position relative to proposed vertiport 20 (JRB).**

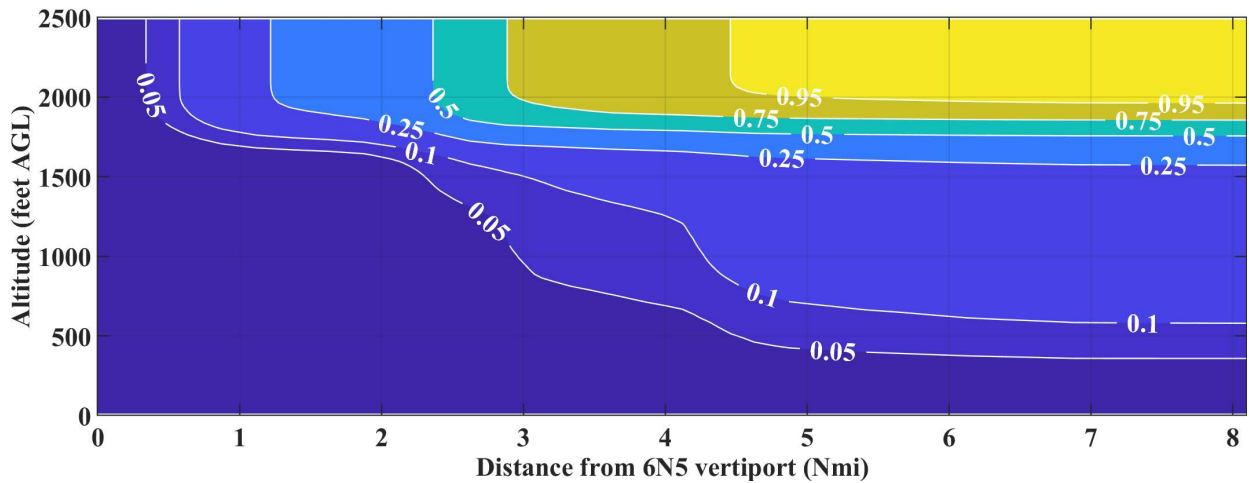

**Figure 18. Rotorcraft position relative to proposed vertiport 21 (6N5).**

Lastly, Figure 19 illustrates the rotorcraft position from vertiport 28 (EWR), which is located away from Manhattan and tourist features of interest. Since the commercial use case of the rotorcraft rarely has the aircraft fly away from the features of interest, Figure 19 suggests the UAM aircraft operating near vertiport 28 will rarely encounter this particular rotorcraft.

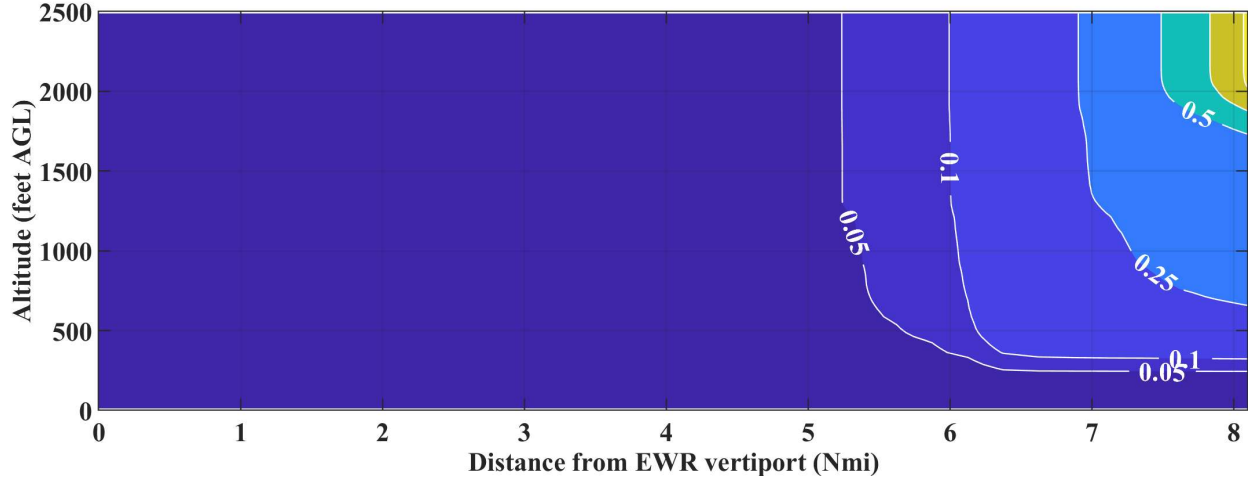

**Figure 19. Rotorcraft position relative to proposed vertiport 28 (EWR).**

This analysis and calculated distributions highlight potential challenges and considerations when designing a UAM network that can interact with other aircraft. This would point to the need for vertiport classes. This analysis is based on three representative aircraft and is insufficient to draw any strong conclusions. Rather we demonstrate an analysis approach that we seek to build upon in future work that would consider many aircraft operating in the airspace.

## V. Conclusion

In this paper we describe a simulation framework developed to assess the impacts of UAM operations on its surrounding airspace. This simulation framework was designed to be modular to allow replacement of any module and isolate the impacts of its assumptions on the system at large. The modular framework also allows for parallel simulation using the MIT Lincoln Laboratory Super Computing Center [24]. Unlike previous studies, our analysis shows the addition of vertiport parking spaces did not correlate as strongly to a reduction in trip delay or average passenger delay in a resource constrained environment. The paper shows the four-passenger capacity configuration, which industry is converging on, will require a large fleet size to meet five percent of the demand for-hire ground vehicles. The fleet size required to meet this level of demand, with average trip delays below 15 minutes, would more than double the current airspace traffic and may impose additional delay in certain high use areas. The analysis also shows, due to the on-demand nature of UAM, not all vertiports will see the same throughput requirements and may justify a need for vertiport class types, where accessing specific vertiports will require an assurance of vehicle turnaround times, vehicle surveillance capabilities, or a fairness slot mechanism. As shown, during planning stage of vertiport placement operators will need to assess the surrounding traffic pattern to understand the airspace risk and interoperability of the respective UAM operations with existing traffic.

## VI. Future Work

UAM operations are likely to suffer from low altitude weather events (e.g., convective weather, winds, low visibility) and weather events should be considered when modeling traffic. Under development, is the integration of weather in the simulation framework and its impacts on the vehicles, not only to calculate weather contingent arrival times, but to assess the feasibility of operations at large. The weather impacts on UAM are expected to not only reduce throughput in network sectors due to separation requirements under low visibility but may cause a complete shutdown of a sector or the entire corridor network. Second, a strict separation criterion is under development where the framework accounts for aircraft sector capacity and allows a strict number of aircraft to operate within the corridor. In the case of weather, sector shutdowns, and separation capacity resource constraints, the framework would need to accommodate alternative route choice selection that would allow traffic to continue to their destination. Futhermore, generalize energy use module will be integrated with each vehicle to understand the impacts of heterogenous vehicle recharge rates on the viability of operations.

## VII. Acknowledgments

This material is based upon work supported by the United States Air Force under Air Force Contract No. FA8702-15-D-0001. Any opinions, findings, conclusions or recommendations expressed in this material are those of the author(s) and do not necessarily reflect the views of the United States Air Force.